\documentclass[11pt]{article}

\usepackage[margin=0.88in]{geometry}
\usepackage[T1]{fontenc}
\usepackage[utf8]{inputenc}
\usepackage{lmodern}
\usepackage{microtype}
\usepackage{xspace}
\usepackage{amsmath,amssymb,amsthm,mathtools}
\usepackage{booktabs,tabularx,array,multirow}
\usepackage{graphicx}
\usepackage{algorithm}
\usepackage{algpseudocode}
\usepackage{enumitem}
\usepackage{xcolor}
\usepackage{natbib}
\usepackage{url}
\usepackage{hyperref}
\usepackage[nameinlink,capitalise,noabbrev]{cleveref}

\hypersetup{
  colorlinks=true,
  linkcolor=blue!55!black,
  citecolor=blue!55!black,
  urlcolor=blue!55!black,
  pdftitle={Prediction-Assisted Pricing and Admission for LLM APIs with Stochastic Token Consumption},
  pdfauthor={Anonymous Authors}
}

\newtheorem{theorem}{Theorem}
\newtheorem{lemma}{Lemma}
\newtheorem{proposition}{Proposition}
\newtheorem{corollary}{Corollary}
\newtheorem{assumption}{Assumption}
\theoremstyle{definition}

\theoremstyle{remark}

\newcommand{\R}{\mathbb{R}}
\newcommand{\E}{\mathbb{E}}
\newcommand{\Prob}{\mathbb{P}}
\newcommand{\one}{\mathbf{1}}
\newcommand{\cA}{\mathcal{A}}

\newcommand{\cF}{\mathcal{F}}
\newcommand{\cX}{\mathcal{X}}

\newcommand{\OPT}{\mathrm{OPT}}
\newcommand{\Reg}{\mathrm{Reg}}
\newcommand{\PCUCB}{\textnormal{\textsc{PC-UCB}}\xspace}

\newcommand{\diam}{\operatorname{diam}}

\title{\textbf{Prediction-Assisted Pricing and Admission for LLM APIs}\\
\textbf{with Stochastic Token Consumption}}
\author{Patrick Wong}
\date{}

\begin{document}
\maketitle

\begin{abstract}
An LLM application often sells or internally allocates several service products: a small or premium model, a short or long token cap, and possibly multiple posted prices. The operational decision is not merely which model answers a prompt. A price changes purchase probability, a token cap changes both user value and the tail of resource consumption, and accepted requests compete for shared compute and premium-model capacity. Demand and output length are initially uncertain, while an offline model may provide useful but imperfect predictions.

We formulate sequential pricing and admission with stochastic resource consumption. Each arriving request belongs to an observable segment. The platform chooses a product--price pair or makes no offer; purchase, revenue, and resource use are then random. An offline predictor supplies a uniform, validated error radius for every segment--product cell. We propose Prediction-Clipped UCB (\PCUCB), which intersects the offline prediction interval with an online confidence interval, evaluates products using resource shadow prices, and reserves a sample-path envelope before commitment. The prior gives a fast start when accurate, while online learning protects the platform when predictions are coarse.

The analysis is modular. On a simultaneous confidence event, regret against a buffered fluid benchmark is bounded by a pacing term plus the cumulative diameter of the intersected intervals. For $J$ segment--product cells and prediction radius $\varepsilon$, this yields
\[
  \widetilde O\!\left(
  \sqrt{T}+(1+\bar\Lambda)
  \min\{T\varepsilon,\sqrt{JT}\}
  \right),
\]
where $\bar\Lambda$ bounds operational shadow prices. Thus the algorithm smoothly interpolates between an almost full-information regime and learning from scratch. Hard feasibility holds on every sample path through reservation envelopes.

Under a large but valid prediction-error radius, \PCUCB obtains 94.8\% of oracle revenue, compared with 90.8\% for a controller that never corrects its predictions and 82.5\% for online UCB without a prior. The paper concludes with an identification and measurement protocol for real LLM API logs, including randomized menu experiments, censored token demand, and service-quality safeguards.
\end{abstract}

\section{Introduction}\label{sec:intro}

Commercial and internal LLM platforms increasingly offer differentiated service rather than a single undifferentiated endpoint. A request can be processed by a small model, a premium model, a retrieval-augmented pipeline, or a tool-using agent. The platform may also choose a maximum output length, latency class, service-level guarantee, or posted price. These choices jointly determine user value and system load. A premium long-context response may command a higher price, but it also creates a heavier and more variable token workload. Selling too many such requests early can crowd out later requests with higher willingness to pay.

This creates an online revenue-management problem with a feature that is especially important for generative models: resource consumption is not known when the offer is made. Output length depends on the prompt, sampling settings, stopping behavior, and whether the user accepts the offer. A service can reserve a worst-case token envelope, but planning only with worst cases is unnecessarily conservative. It can plan using expected consumption, but then it needs a separate mechanism to prevent a rare long response from violating physical capacity. The correct architecture therefore separates \emph{economic planning} from \emph{sample-path admission}.

Learning is equally central. A new product may have no reliable demand curve. Even for an existing product, a model update can change completion quality, user acceptance, and token length. Yet platforms often possess an offline predictor built from benchmark scores, historical cohorts, survey data, or a launch experiment. Such a predictor should not be discarded. Nor should it be trusted indefinitely. What is needed is a controller that exploits a validated prediction while retaining the ability to learn away its residual error.

We study this problem through a finite product menu. Each action combines a model tier, a token cap, and a price. A request arrives with an observable segment; the platform chooses one action or makes no offer. If the request purchases, the platform receives the posted price and observes a random resource vector. The resource coordinates can represent expected GPU-seconds, premium-model capacity, cached-context bandwidth, external API spend, or a service-level budget. This discrete menu is already rich enough to expose the interaction between pricing, model allocation, and token reservation. Continuous prices can be handled by discretization or a demand model, as discussed later.

Our algorithm, Prediction-Clipped UCB (\PCUCB), combines three ideas. First, the offline model gives an interval rather than a point estimate. Second, online observations give a conventional statistical confidence interval. Their intersection is never wider than either source alone. Third, a vector of shadow prices converts expected resource use into an internal marginal cost. The controller offers the product with the largest confidence-adjusted revenue net of resource charges, subject to a reservation meter. A projected dual update raises the internal price of an overused resource and lowers it when capacity is underused.

\subsection{Contributions}

\paragraph{Joint pricing, product selection, and admission.}
We formulate an LLM service in which each action specifies a price, model tier, and token cap. Purchase probability, revenue, and conditional token use can all depend on the request segment. The model allows multiple capacities and an explicit no-offer action. A reservation envelope guarantees that an indivisible accepted job never pushes the system beyond physical capacity.

\paragraph{A prediction-assisted online controller.}
\PCUCB intersects a validated offline interval with an online confidence interval. It uses an upper confidence estimate for revenue and a conservative upper estimate for resource consumption. This choice encourages exploration only when the prospective revenue justifies both statistical uncertainty and current scarcity. The controller reduces to a full-information shadow-price policy when the predictor is exact and to a standard online-learning controller when the prior radius is uninformative.

\paragraph{An interpolation guarantee.}
The main theorem separates the regret of the allocation controller from the quality of the estimator. A counting lemma bounds the sum of cell-specific interval widths by
$O(\min\{T\varepsilon,\sqrt{JT\log T}\})$. The resulting guarantee quantifies the operational value of better predictions and avoids a discontinuous choice between ``use the forecast'' and ``ignore the forecast.''

\paragraph{A reproducible stress test and a real-data protocol.}
The attached code generates the complete synthetic data, table, and figures. The experiment is designed to reveal both the benefit and the cost of online correction: when errors are small, prediction-only control can be marginally better because exploration is not free; when errors are large, clipping with online evidence materially improves robustness. We then specify what must be randomized, logged, and audited in a real deployment.

\subsection{What the paper does not claim}

The numerical study is synthetic and should not be read as evidence about any named model provider, user population, or market price. The theoretical model assumes a valid error envelope for the offline predictor, stationary cell means over the analyzed horizon, and bounded resource envelopes. It also treats users as one-shot arrivals. Strategic delay, repeated users, fairness constraints, and queueing-dependent latency require additional modeling. These limitations are explicit because a pricing policy can affect users directly and should not be deployed solely on the basis of an offline simulation.

\section{Related Work}\label{sec:related}

\subsection{Revenue management and dynamic pricing}

Classical revenue management studies how to accept, reject, and price demand against finite inventory \citep{gallego1994optimal,talluri2004revenue}. Network formulations use a deterministic or stochastic linear program to convert shared capacities into bid prices. Re-solving and self-adjusting controls update those prices as demand and inventory evolve \citep{jasin2014reoptimization}. When demand is unknown, the controller must balance learning with capacity consumption; examples include dynamic pricing with unknown demand and Thompson-sampling approaches to network revenue management \citep{besbes2009dynamic,ferreira2018online}.

The information structure closest to ours is the setting in which a machine-learned price or demand estimate arrives with a certified error bound. \citet{ao2026learning} study how such information changes attainable regret in resource-constrained pricing. Our action is a bundled LLM service product rather than a scalar price, and our observation includes stochastic token use in addition to purchase. We use an interval-intersection controller whose proof exposes a generic confidence-diameter term.
Many sharp revenue-management results impose regularity or nondegeneracy to keep the fluid solution stable. In discrete menus, degeneracy is natural: two prices may have almost identical adjusted revenue, or a premium-capacity constraint may become binding at the same point as compute. \citet{jiang2025degeneracy} show that logarithmic regret can be achieved in an important network setting without conventional nondegeneracy assumptions. Our more general learning bound is slower, but likewise avoids assuming a unique basic solution. We compare actions through their current Lagrangian scores and use only bounded dual prices.

\subsection{Online allocation and knapsack learning}

Bandits with knapsacks combine reward learning with irreversible resource consumption \citep{badanidiyuru2013bandits}; convex-knapsack and dual methods broaden the action and objective classes \citep{agrawal2014bandits,balseiro2023best}. These models clarify why ordinary regret against the best fixed arm is insufficient: an apparently profitable action can be undesirable because it consumes a resource needed by future arrivals. Our controller uses the same economic principle but exploits a product-level prediction interval and enforces an explicit per-job reservation envelope.

Online stochastic knapsack and prophet inequalities study what can be achieved against offline or LP benchmarks under random arrivals. The best achievable ratio depends on item size, information, and the chosen relaxation. The sample-path utilization arguments in \citet{jiang2025tight} are particularly relevant to stochastic token demand. We do not claim their tight approximation factor in our multi-resource learning model. Instead, we use their distinction as design guidance: mean resource use determines the bid-price decision, while a hard reservation check protects feasibility when an individual completion is long.

\subsection{LLM routing and inference systems}

LLM routing chooses among models or cascades to reduce cost while retaining quality \citep{chen2023frugalgpt,ong2024routellm,hu2024routerbench}. Most routing benchmarks treat model prices as exogenous and evaluate requests independently. Our focus is complementary: the service chooses an offered product and price, and all accepted requests share workload-level capacities. Inference systems such as Orca, vLLM, and FlexGen demonstrate that batching, memory management, and offloading shape the feasible throughput frontier \citep{yu2022orca,kwon2023vllm,sheng2023flexgen}. We abstract their measurements into resource vectors and reservation envelopes. This lets the economic controller react to current scarcity without pretending to replace a lower-level scheduler.

\section{System Model}\label{sec:model}

\subsection{Arrivals, segments, and products}

Time is indexed by $t=1,\ldots,T$. A request arrives with an observed segment $X_t\in\cX=\{1,\ldots,K\}$. A segment can combine task type, organization tier, latency sensitivity, predicted prompt length, and other pre-offer features. We assume $X_t$ is observed before the product is selected and that the sequence is i.i.d. with probabilities $q_x$ in the baseline analysis. The controller may estimate $q_x$ online; this adds a conventional concentration term and is omitted from the main notation.

The finite menu is $\cA=\{1,\ldots,A\}$. Product $a$ specifies at least
\[
  a=(m_a,\ell_a,p_a),
\]
where $m_a$ is a model tier, $\ell_a$ is a token or compute cap, and $p_a$ is the posted price. It may also include retrieval, tool access, or a latency class. Action $0$ denotes no offer and has zero reward and zero resource use.

After action $A_t=a$ is offered, a purchase indicator $D_t(a)\in\{0,1\}$ is realized. Revenue is
\[
  R_t(a)=p_aD_t(a)\in[0,1],
\]
after normalization. The resource vector is $C_t(a)\in[0,1]^m$ and is zero when there is no purchase. Conditional on purchase, it can include stochastic output tokens, GPU time, premium-model occupancy, memory-time, or monetary vendor spend. We write
\[
  r_{x,a}=\E[R_t(a)\mid X_t=x],
  \qquad
  c_{x,a}=\E[C_t(a)\mid X_t=x].
\]
The pair $(r_{x,a},c_{x,a})$ is unknown.

\subsection{Capacity and reservation envelopes}

The workload has capacity $B=Tb\in\R_+^m$. The policy must satisfy
\begin{equation}
  \sum_{t=1}^T C_t(A_t)\le Tb
  \quad\text{componentwise on every sample path.}
  \label{eq:hard-feasible}
\end{equation}
For every product, the service knows an envelope $u_a\in[0,1]^m$ such that
\begin{equation}
  C_t(a)\le u_a \quad\text{almost surely}.
  \label{eq:envelope}
\end{equation}
A token cap supplies a natural envelope for token-related resources. For latency or GPU time, the envelope can be a conservative admission reservation supplied by the scheduler. Before offering $a$, the controller checks whether $u_a$ fits the remaining capacity. This can be conservative near the end of the horizon, but it eliminates the ambiguity of ``expected feasibility.''

\subsection{Offline predictions with a valid radius}

An offline system provides predictions $(\widetilde r_{x,a},\widetilde c_{x,a})$. We assume the model card or validation procedure supplies a simultaneous radius $\varepsilon\in[0,1]$ such that
\begin{equation}
  |\widetilde r_{x,a}-r_{x,a}|\le\varepsilon,
  \qquad
  \|\widetilde c_{x,a}-c_{x,a}\|_\infty\le\varepsilon
  \quad\text{for all }(x,a).
  \label{eq:prior-bound}
\end{equation}
Cell-specific radii are allowed, but a common radius keeps the theorem readable. The bound is an assumption about validation, not a claim that an arbitrary neural predictor is calibrated. In a real study it should be estimated on a held-out, launch-representative sample with multiplicity correction.

The point of \eqref{eq:prior-bound} is not that the forecast is perfect. Rather, it gives a finite region in which online learning must search. When $\varepsilon=0$, the means are known. When $\varepsilon=1$, the interval carries almost no information and the algorithm behaves like a conventional bandit controller.

\subsection{Feedback}

After an offered product is accepted or rejected, the controller observes revenue and realized resource use for that product. No counterfactual purchase outcome is observed for other products. This is ordinary bandit feedback. We assume the product itself can be logged before purchase and that a rejection is an observed zero-revenue, zero-resource outcome. If users can abandon without a reliable exposure log, then the observed data are selectively missing and the confidence intervals below are invalid without an additional observation model.

\section{Fluid Benchmark and Shadow Prices}\label{sec:benchmark}

\subsection{Deterministic linear relaxation}

Let $y_{x,a}$ denote the probability of offering product $a$ to segment $x$. The fluid relaxation is
\begin{align}
  \OPT(b)=\max_{y\ge0}\quad
  &T\sum_{x\in\cX}q_x\sum_{a\in\cA}y_{x,a}r_{x,a}\label{eq:fluid}\\
  \text{s.t.}\quad
  &\sum_{x\in\cX}q_x\sum_{a\in\cA}y_{x,a}c_{x,a}\le b,\nonumber\\
  &\sum_{a\in\cA}y_{x,a}\le1,\qquad x\in\cX.\nonumber
\end{align}
The missing probability is the no-offer action. This benchmark knows the mean demand and mean resource use, but not individual future outcomes. It is an upper bound on the expected revenue of policies that only have aggregate expected-capacity constraints. With hard envelopes and indivisible jobs, the relationship to the exact offline optimum can depend on item size; we therefore state our principal guarantee against a buffered version of \eqref{eq:fluid}.

For a dual price $\lambda\in\R_+^m$, the Lagrangian score of cell $(x,a)$ is
\begin{equation}
  s_{x,a}(\lambda)=r_{x,a}-\lambda^\top c_{x,a}.
  \label{eq:true-score}
\end{equation}
For fixed $\lambda$, each segment chooses a product with maximum positive score. The resource prices coordinate otherwise independent segment decisions.

\subsection{Buffered benchmark}

Let $b'=b-\gamma\one$ for a buffer $\gamma>0$ satisfying $b'>0$. We compare the online policy to $\OPT(b')$ and account separately for the loss from the buffer. The next assumption is a local value-sensitivity condition.

\begin{assumption}[Bounded shadow prices]\label{ass:dual}
For every rate on the segment between $b'$ and $b$, the fluid LP admits an optimal dual vector with $\ell_1$ norm at most $\Lambda_\star$.
\end{assumption}

Under \Cref{ass:dual}, concavity of the fluid value implies
\begin{equation}
  \OPT(b)-\OPT(b')\le T\Lambda_\star m\gamma.
  \label{eq:buffer-loss}
\end{equation}
This is the opportunity cost of holding back capacity for stochastic deviations and final-job envelopes.

\subsection{Why a unique fluid solution is unnecessary}

The set of score maximizers may contain several products. Our algorithm uses an arbitrary deterministic tie breaker and projects prices onto a bounded box. The analysis compares the selected score to the best score under the same current price; it never differentiates an optimal basis or assumes that a small perturbation leaves the primal solution unchanged. Ties can change the product mix, but they do not invalidate the one-step Lagrangian comparison. This is important for menus in which adjacent prices or caps are intentionally close substitutes.

\section{Prediction-Clipped UCB}\label{sec:algorithm}

\subsection{Online confidence intervals}

Index a segment--product cell by $j=(x,a)$ and let $J=KA$. Before round $t$, let $N_j(t)$ be the number of times cell $j$ has been offered, and let $\widehat r_j(t)$ and $\widehat c_j(t)$ be its sample means. Define
\begin{equation}
  \alpha_t(n)=\sqrt{\frac{2\log(2J(m+1)T/\delta)}{\max\{1,n\}}}.
  \label{eq:online-radius}
\end{equation}
We clip all endpoints to $[0,1]$. The online confidence intervals are
\begin{align}
  I^r_j(t)&=[\widehat r_j(t)-\alpha_t(N_j(t)),\widehat r_j(t)+\alpha_t(N_j(t))],\nonumber\\
  I^c_{j,i}(t)&=[\widehat c_{j,i}(t)-\alpha_t(N_j(t)),\widehat c_{j,i}(t)+\alpha_t(N_j(t))].
\end{align}
The prediction intervals are
\begin{align}
  P^r_j&=[\widetilde r_j-\varepsilon,\widetilde r_j+\varepsilon],\nonumber\\
  P^c_{j,i}&=[\widetilde c_{j,i}-\varepsilon,\widetilde c_{j,i}+\varepsilon].
\end{align}
The algorithm intersects the two sources:
\begin{equation}
  H^r_j(t)=I^r_j(t)\cap P^r_j,
  \qquad
  H^c_{j,i}(t)=I^c_{j,i}(t)\cap P^c_{j,i}.
  \label{eq:intersection}
\end{equation}
On the simultaneous confidence event, every intersection is nonempty because both component intervals contain the true mean. In implementation, an empty intersection is a diagnostic that the advertised prediction radius or statistical model is invalid. The controller can then replace the intersection by the convex hull and raise a monitoring alert.

Let $U^r_j(t)$ be the upper endpoint of $H^r_j(t)$ and $U^c_{j,i}(t)$ the upper endpoint of $H^c_{j,i}(t)$. The first is optimistic about revenue; the second is conservative about consumption. Define the effective diameter
\begin{equation}
  w_j(t)=\max\left\{\diam H^r_j(t),\max_i\diam H^c_{j,i}(t)\right\}.
  \label{eq:effective-width}
\end{equation}
Because $H$ is an intersection,
\begin{equation}
  w_j(t)\le2\min\{\varepsilon,\alpha_t(N_j(t))\}.
  \label{eq:width-min}
\end{equation}

\subsection{Confidence-adjusted economic score}

The controller maintains a vector $\lambda_t\in[0,\bar\Lambda]^m$ of resource shadow prices. For segment $x$ and product $a$, it computes
\begin{equation}
  \widehat s_{x,a,t}
  =U^r_{x,a}(t)-\lambda_t^\top U^c_{x,a}(t).
  \label{eq:estimated-score}
\end{equation}
It ranks products by this score, keeps only products with positive score, and then applies the reservation meter. Using an upper rather than lower consumption endpoint is a deliberate safety adjustment. It may underexplore an uncertain resource-intensive product, but it avoids treating uncertain token demand as free.

\subsection{Dual update}

After the chosen action's realized resource vector $C_t$ is observed, prices update as
\begin{equation}
  \lambda_{t+1}
  =\Pi_{[0,\bar\Lambda]^m}\left[
      \lambda_t+\eta(C_t-b')
    \right],
  \label{eq:dual-update}
\end{equation}
where $\Pi$ is Euclidean projection. This is a virtual-queue or online-gradient update. A resource used faster than its target rate becomes more expensive. The update uses realized consumption so it automatically reacts to output-length shocks.

\begin{algorithm}[t]
\caption{Prediction-Clipped UCB (\PCUCB)}\label{alg:pcucb}
\begin{algorithmic}[1]
\Require Horizon $T$, capacity $Tb$, buffer $\gamma$, envelopes $u_a$, prediction intervals, step size $\eta$, price cap $\bar\Lambda$.
\State Set $b'=b-\gamma\one$, $\lambda_1=0$, remaining capacity $B_1=Tb$, and zero cell counts.
\For{$t=1,\ldots,T$}
  \State Observe segment $X_t=x$.
  \For{each product $a\in\cA$}
    \State Intersect its online and prediction intervals as in \eqref{eq:intersection}.
    \State Compute the score \eqref{eq:estimated-score}.
  \EndFor
  \State Rank products by score and append the no-offer action.
  \State Choose the highest-ranked positive-score product satisfying $u_a\le B_t$; choose no offer if none exists.
  \State Post the price and execute the accepted job, obtaining $R_t$ and $C_t$.
  \State Set $B_{t+1}=B_t-C_t$ and update the selected cell's statistics.
  \State Update $\lambda_{t+1}$ by \eqref{eq:dual-update}.
\EndFor
\end{algorithmic}
\end{algorithm}

\subsection{Operational variants}

The algorithm can be implemented with a precomputed menu table. For each segment, product, and price bucket, the service stores the prediction interval, online sufficient statistics, and reservation envelope. The per-request optimization is then a scan over products. When the menu is large, products can be screened by dominance: if one product has no larger revenue upper bound and no smaller consumption lower bound than another, it need not be offered at the current update.

The algorithm also permits a separate premium-model admission layer. In that architecture, \PCUCB chooses a product using expected resource use, while the scheduler can reject or downgrade the request if instantaneous queueing conditions make the reservation unavailable. Such interventions should be logged as administrative censoring rather than user rejection.

\section{Theoretical Analysis}\label{sec:theory}

\subsection{Assumptions and regret}

We analyze stationary cell means and i.i.d. segments. Outcomes are conditionally independent across time given the selected cells and lie in $[0,1]^{m+1}$. The prediction bound \eqref{eq:prior-bound} is valid. Reservation envelopes satisfy \eqref{eq:envelope}. The shadow-price cap obeys $\bar\Lambda\ge\Lambda_\star$.

Let $V_T^{\PCUCB}$ be the expected revenue of \Cref{alg:pcucb}. We define regret against the unbuffered fluid benchmark as
\[
  \Reg_T=\OPT(b)-V_T^{\PCUCB}.
\]
The theorem decomposes this quantity into the price-learning term, confidence-diameter term, buffer loss, and a final reservation correction.

\subsection{Confidence event}

\begin{lemma}[Simultaneous cell confidence]\label{lem:confidence}
With probability at least $1-\delta$, every true reward and resource mean belongs to its online interval for all cells and all times. On the same event, every intersection in \eqref{eq:intersection} contains the true mean and satisfies \eqref{eq:width-min}.
\end{lemma}

The proof is a union bound over cells, output coordinates, and sample sizes, using a bounded-difference concentration inequality. It appears in \Cref{app:proof-confidence}.

\begin{lemma}[One-step score comparison]\label{lem:score}
On the event of \Cref{lem:confidence}, for any $\lambda\in[0,\bar\Lambda]^m$ and any cell $j$,
\[
  \left|
  \big(U^r_j-\lambda^\top U^c_j\big)
  -\big(r_j-\lambda^\top c_j\big)
  \right|
  \le(1+\bar\Lambda)w_j.
\]
If the reservation meter does not override the score maximizer at round $t$, the selected product $A_t$ satisfies
\[
  s_{X_t,A_t}(\lambda_t)
  \ge
  \max\left\{0,\max_{a\in\cA}s_{X_t,a}(\lambda_t)\right\}
  -2(1+\bar\Lambda)w_{X_t,A_t}^{\max}(t),
\]
where $w_{X_t,A_t}^{\max}(t)$ can be replaced by the maximum width among the selected and true score-maximizing products.
\end{lemma}

The first inequality follows because both the chosen endpoint and true mean lie in an interval of diameter $w_j$. The second is the standard ``estimated maximizer versus true maximizer'' argument.

\subsection{A confidence-to-control theorem}

Let $M_T$ denote the expected number of rounds on which the reservation meter overrides the unconstrained score maximizer. Since reward is at most one, its contribution to regret is at most $M_T$.

\begin{theorem}[Confidence-to-control bound]\label{thm:generic}
Suppose \Cref{ass:dual} holds and the algorithm uses $\eta=\bar\Lambda/\sqrt{mT}$. On the event of \Cref{lem:confidence},
\begin{align}
  \Reg_T
  \le{}&
  2(1+\bar\Lambda)
  \E\left[\sum_{t=1}^T \overline w_t\right]
  +2\bar\Lambda\sqrt{mT}
  +T\Lambda_\star m\gamma
  +M_T
  +\delta T,
  \label{eq:generic-bound}
\end{align}
where $\overline w_t$ is the larger width of the algorithm's selected cell and a true score-maximizing cell under $\lambda_t$.
\end{theorem}

The proof in \Cref{app:proof-generic} combines \Cref{lem:score} with online projected-gradient regret for the dual sequence. The statement intentionally does not require a unique fluid optimizer. The term $\delta T$ covers the complement of the confidence event. A refined analysis can replace it by a problem-dependent failure loss.

The width of an unselected score-maximizing cell raises a familiar exploration issue. One can guarantee that it is sampled by forcing each segment--product cell a logarithmic number of times, or by using optimistic scores as in \PCUCB. The next counting result captures the cumulative width of visited cells; the standard UCB charging argument extends it to the comparator cells up to constants.

\begin{lemma}[Prediction-assisted width sum]\label{lem:width-sum}
Let $j_t\in\{1,\ldots,J\}$ be any sequence of visited cells and let $N_j(t)$ be its prior visit count. For $L\ge1$,
\[
  \sum_{t=1}^T
  \min\left\{\varepsilon,\sqrt{\frac{L}{\max\{1,N_{j_t}(t)\}}}\right\}
  \le
  \min\left\{T\varepsilon,\;2\sqrt{LJT}+J\sqrt{L}\right\}.
\]
\end{lemma}

The first branch follows by bounding every term by $\varepsilon$. The second sums $1/\sqrt n$ within each cell and applies Cauchy--Schwarz. The additive $J\sqrt L$ covers the first observation of each cell.

\begin{theorem}[Prediction-to-learning interpolation]\label{thm:main}
Under the assumptions above, choose
\[
  \gamma=c_0\sqrt{\frac{\log(mT/\delta)}{T}}
\]
large enough to cover the resource martingale and virtual-queue deviation. Then the expected regret of \PCUCB satisfies
\begin{align}
  \Reg_T
  =\widetilde O\left(
    \sqrt{T}
    +(1+\bar\Lambda)
      \min\{T\varepsilon,\sqrt{JT}\}
    +M_T
  \right),
  \label{eq:main-bound}
\end{align}
where logarithmic factors depend on $J,m,T$, and $1/\delta$. If each envelope is $o(T)$ relative to total capacity and the buffer dominates stochastic deviations, then $M_T=O(1)$ with high probability; regardless of this event, the policy remains hard feasible.
\end{theorem}

\begin{corollary}[Two information regimes]\label{cor:regimes}
Ignoring logarithmic factors and final-meter corrections:
\begin{enumerate}[leftmargin=1.6em]
  \item If $\varepsilon\lesssim 1/\sqrt T$, prediction error contributes at most $O(\sqrt T)$ and the controller has the same first-order rate as a known-model policy.
  \item If the prediction is uninformative, the bound becomes $O((1+\bar\Lambda)\sqrt{JT})$, matching the cell-learning scale up to logarithmic and allocation terms.
\end{enumerate}
\end{corollary}

The guarantee makes the value of a more accurate forecast explicit. Reducing $\varepsilon$ matters linearly until online sampling becomes more informative than the prior. Beyond that point, further offline improvement has little effect on the worst-case statistical term, although it can still improve constants and early-horizon behavior.

\subsection{Hard feasibility}

\begin{proposition}[Pathwise feasibility]\label{prop:feasibility}
If the policy offers product $a$ only when $u_a$ is componentwise no larger than remaining capacity and realized use satisfies \eqref{eq:envelope}, then \eqref{eq:hard-feasible} holds for every outcome path.
\end{proposition}

This proposition is elementary but operationally important. It does not require concentration, a correct forecast, or a stable demand distribution. The price-learning theorem determines how much value is lost relative to a fluid benchmark; the meter determines whether a job may physically start.

\subsection{Misspecified prediction radii}

A valid prediction interval is the cleanest information model, but it can fail after a distribution shift. Let
\[
  \zeta_j=	ext{distance of the true cell-mean vector from its advertised prediction box}.
\]
If intersections are replaced by convex hulls when empty, the proof adds an approximation term of order
\[
  (1+\bar\Lambda)\sum_{t=1}^T\zeta_{j_t}.
\]
This observation suggests a monitoring rule: persistent empty intersections should increase the advertised radius or trigger a product-specific reset. Quietly retaining an invalid narrow box can create linear regret because online evidence is clipped away.

\subsection{Discussion of degeneracy}

The theorem is intentionally not logarithmic. It allows many cells, unknown rewards and costs, and only a coarse prediction radius. Stronger rates are possible under structural conditions on demand, second-order growth, or stable dual solutions. The contribution here is the interpolation form and the hard-feasible LLM product model. In particular, the proof does not use a ``margin'' between the best and second-best product. If two products are tied, selecting either has negligible score regret, though their resource mix can influence the dual trajectory.

\section{Structural Implications for LLM Product Design}\label{sec:structure}

\subsection{The economically relevant quantity is not accuracy per dollar}

For segment $x$, product $a$ is attractive when
\[
  r_{x,a}-\lambda^\top c_{x,a}>0.
\]
The first term already includes acceptance probability and price. The second prices every scarce operational input. A premium model can therefore be optimal even when its raw revenue per expected token is lower, provided it uses a different scarce resource mix or serves a segment with sufficiently high willingness to pay. Conversely, a model with excellent benchmark accuracy may be rejected if its token tail consumes a resource with a high current shadow price.

This is different from ranking models by a static ``quality divided by cost'' score. A ratio has no consistent meaning with multiple resources, and it ignores the outside option. Shadow prices provide the correct local exchange rate between revenue and each capacity.

\subsection{Token caps have option value}

A shorter cap affects both demand and the resource envelope. It may lower user value and acceptance, but it allows the platform to admit a job when the long-cap product no longer fits. This makes caps useful even when average output length is far below the long limit. Near capacity exhaustion, the envelope rather than the mean determines which products remain feasible.

\begin{proposition}[Cap substitution]\label{prop:cap}
Consider two products with the same model and price, where product $S$ has a smaller resource envelope than product $L$. If $S$ has positive adjusted mean score and $L$ is infeasible under remaining capacity while $S$ is feasible, then a reservation-aware menu weakly dominates a menu containing only $L$ on that sample path.
\end{proposition}

The statement is immediate because the larger menu can emulate the smaller one and gains a feasible positive-score alternative. The nontrivial empirical question is how much demand a shorter cap loses.

\subsection{When predictions are most valuable}

The width bound identifies three determinants of prediction value. First, a short horizon gives little time to learn, so a prior is especially useful. Second, a large menu or many segments increase $J$ and make from-scratch exploration expensive. Third, a high shadow-price norm magnifies resource-prediction errors. In a highly constrained premium tier, improving token-use calibration may be more valuable than improving purchase-probability calibration by the same absolute amount.

\subsection{Why expected use and reservation should remain separate}

Replacing $c_{x,a}$ by $u_a$ in the economic score guarantees feasibility but prices every job at its worst case. For heavy-tailed output length, this can destroy utilization. Conversely, using only $c_{x,a}$ without a meter can violate capacity on a rare path. The two-layer design uses mean consumption for planning and a bounded envelope for commitment. A lower-level scheduler can make the envelope dynamic, for example by reserving less when preemption or spillover capacity is available.

\section{Synthetic Experiments}\label{sec:experiments}

\subsection{Questions}

The experiment is designed around four questions:
\begin{enumerate}[leftmargin=1.6em]
  \item Does a correct narrow prediction interval provide a meaningful fast start?
  \item When the prediction is biased but remains inside its advertised radius, does online clipping improve robustness?
  \item Can the controller pace two resources without using worst-case envelopes in its economic score?
  \item What failure mode appears when price optimization ignores shared capacity?
\end{enumerate}
The experiment is illustrative rather than calibrated to a specific provider.

\subsection{Data-generating process}

There are $T=6000$ arrivals and three segments with probabilities $(0.46,0.34,0.20)$. The menu contains sixteen products formed by two model tiers, two token caps, and four prices per tier. Purchase probability is logistic in segment value, model-tier value, cap value, and price. Revenue equals price times the purchase indicator.

There are two resources. The first is normalized compute use and varies by segment, tier, cap, purchase, and an idiosyncratic output-length shock. The second is premium-model capacity and is much larger for the premium tier. Per-period capacities are $(0.235,0.115)$. Each product has a deterministic reservation envelope derived from its tier and cap.

Offline predictions are perturbed from the true cell means by a structured error with radius
\[
  \varepsilon\in\{0,0.02,0.05,0.10,0.18\}.
\]
The perturbation intentionally overpredicts revenue and underpredicts resource use for expensive premium, long-cap products, a direction that can be operationally harmful. The entire perturbation is rescaled so the advertised radius remains valid. Results average ten independent arrival and outcome traces. Error bars are 95\% normal intervals across repetitions.

\subsection{Policies}

\paragraph{Oracle.}
Uses the true cell means in the same shadow-price controller and reservation meter. It is not the exact dynamic-program optimum, but it isolates statistical loss.

\paragraph{Prediction-Clipped UCB.}
Uses the intersection controller in \Cref{alg:pcucb}.

\paragraph{Prediction only.}
Uses the offline point prediction throughout and never learns from realized outcomes.

\paragraph{Online UCB.}
Discards the offline prediction and learns cell means from scratch.

\paragraph{Myopic prediction.}
Chooses the highest predicted revenue among products whose envelopes fit, ignoring shadow prices. It exposes the cost of treating each request independently.

\subsection{Main result}

\begin{table}[t]
\centering
\caption{Synthetic stress test at prediction radius $\varepsilon=0.18$. Revenue intervals are mean $\pm$ 95\% confidence half-width across ten repetitions. ``No offer'' counts include rounds in which every positive-score product is screened out or fails the reservation check.}
\label{tab:main-results}
\small
\begin{tabular}{lrrrrr}
\toprule
Policy & Revenue & Oracle share & Compute util. & Premium util. & No offer \\
\midrule
Oracle & 2303.9 $\pm$ 46.1 & 100.0\% & 100.0\% & 99.7\% & 54 \\
Prediction-clipped UCB & 2183.5 $\pm$ 0.2 & 94.8\% & 99.2\% & 100.0\% & 91 \\
Prediction only & 2092.9 $\pm$ 9.6 & 90.8\% & 97.4\% & 100.0\% & 137 \\
Online UCB & 1900.3 $\pm$ 34.2 & 82.5\% & 98.0\% & 100.0\% & 136 \\
Myopic prediction & 949.2 $\pm$ 109.0 & 41.2\% & 55.6\% & 100.0\% & 4051 \\
\bottomrule
\end{tabular}
\end{table}

\Cref{tab:main-results} shows the large-error stress test. \PCUCB reaches 94.8\% of oracle revenue. Prediction-only control reaches 90.8\%, so online correction recovers approximately four percentage points of oracle revenue. Online UCB reaches 82.5\%; it eventually learns useful cells but pays a large cold-start cost over a finite horizon. The myopic controller earns only 41.2\% of oracle revenue. It rapidly saturates premium capacity, leaves much of the general compute capacity unused, and consequently makes no offer on roughly two thirds of arrivals.

The proposed policy uses essentially all premium capacity and 99.2\% of compute capacity. This is a useful diagnostic: the revenue improvement over prediction-only is not produced by simply refusing more demand. It comes from a different product mix under the same scarce premium resource.

\begin{figure}[t]
\centering
\includegraphics[width=0.78\textwidth]{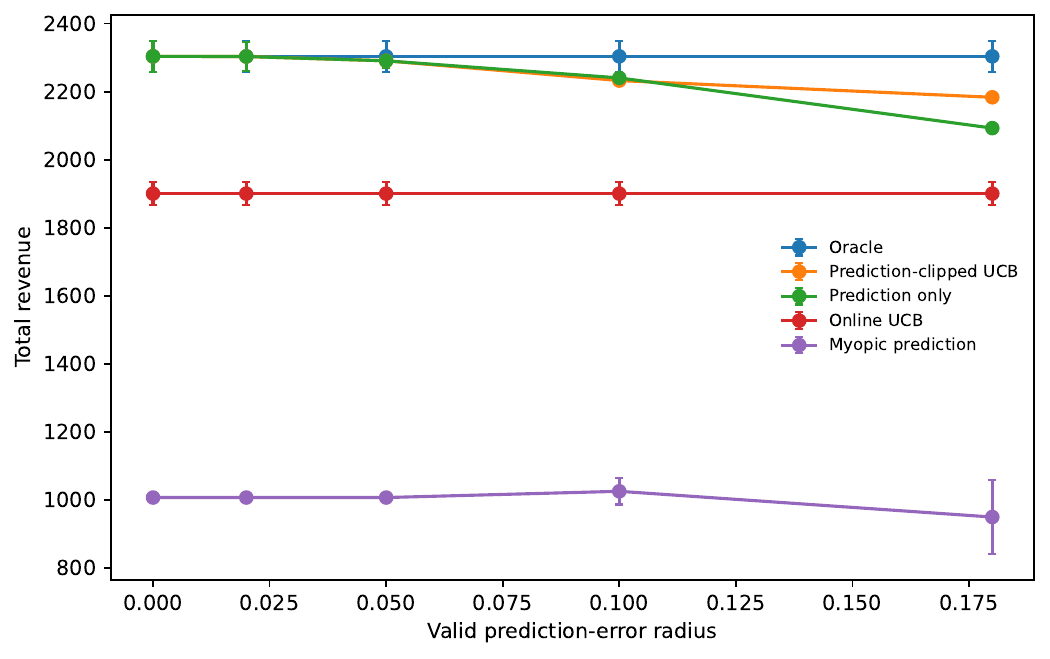}
\caption{Total revenue as the valid offline prediction-error radius varies. Prediction-only control is competitive at small errors, while online clipping becomes valuable as the prior becomes coarse.}
\label{fig:error-revenue}
\end{figure}

\Cref{fig:error-revenue} shows the full error sweep. At $\varepsilon=0$, the oracle, prediction-only policy, and \PCUCB coincide. At $\varepsilon=0.02$, the prediction remains sufficiently accurate that online correction has almost no effect. At $\varepsilon=0.10$, prediction-only is marginally ahead of \PCUCB in this design (97.3\% versus 96.9\% of oracle). That difference is informative rather than embarrassing: exploration and conservative cost uncertainty are not free. The purpose of a hybrid method is robustness across error regimes, not mechanical dominance on every finite sample. At $\varepsilon=0.18$, the online evidence is sufficiently valuable to outweigh its exploration cost.

\begin{figure}[t]
\centering
\includegraphics[width=0.78\textwidth]{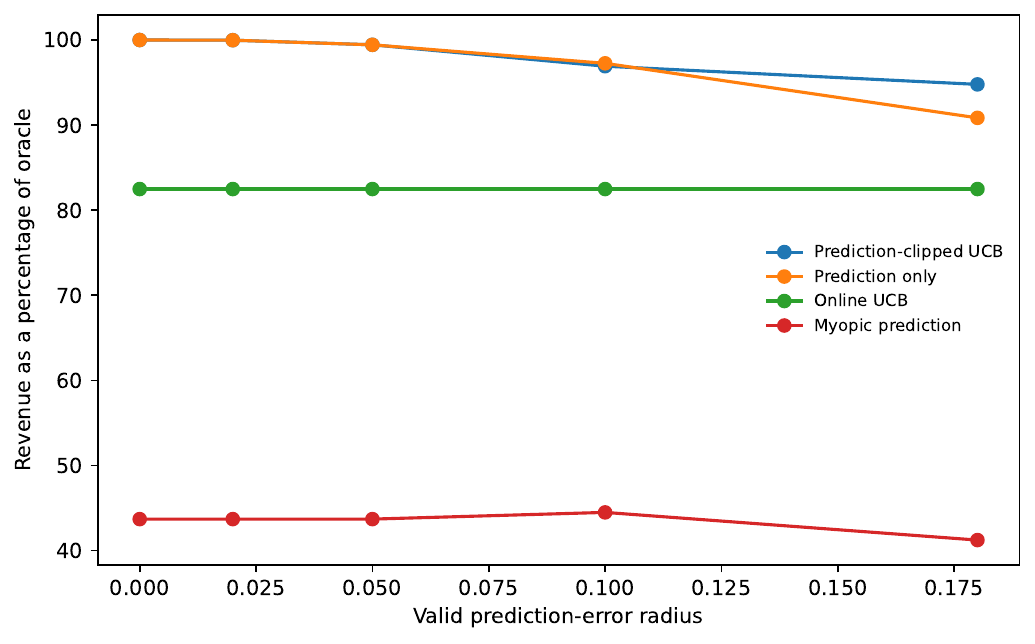}
\caption{Revenue relative to the oracle shadow-price controller. The hybrid policy interpolates between trusting a good prior and learning from data.}
\label{fig:oracle-share}
\end{figure}

\subsection{Within-horizon learning}

\begin{figure}[t]
\centering
\includegraphics[width=0.79\textwidth]{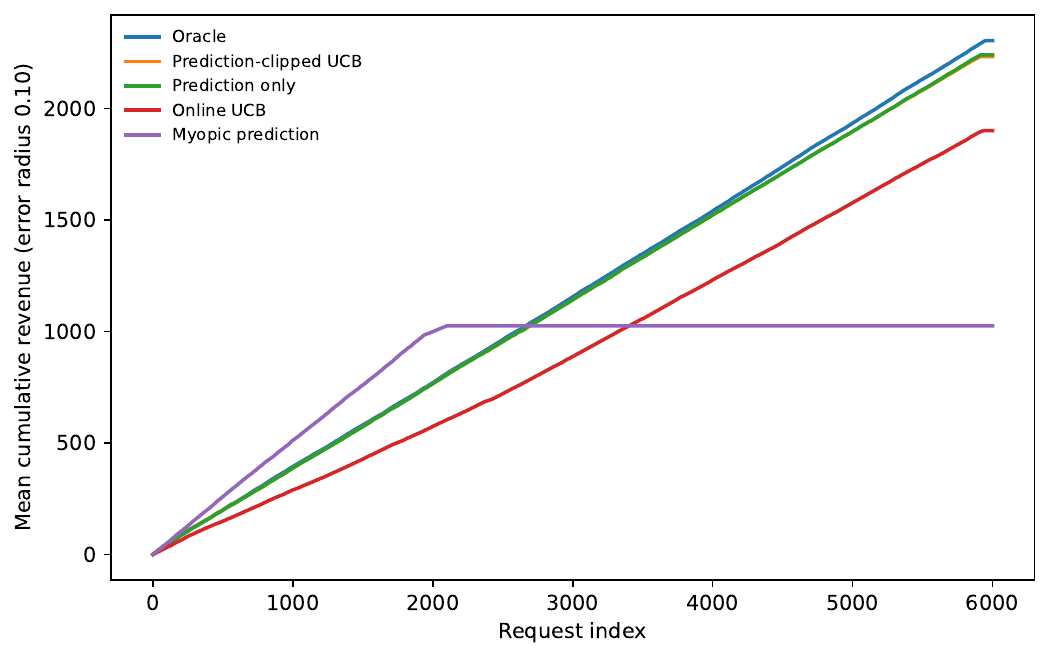}
\caption{Mean cumulative revenue at $\varepsilon=0.10$. The online-only policy loses revenue early because every segment--product cell is initially uncertain.}
\label{fig:cumulative}
\end{figure}

The cumulative curves in \Cref{fig:cumulative} clarify the mechanism. The prediction-based policies begin near the oracle slope. Online UCB initially explores several low-value price--product cells and never fully recovers over 6000 rounds. \PCUCB stays close to prediction-only because the prior interval clips most excessively optimistic online bonuses. Its slight middle-horizon gap comes from conservative upper bounds on resource consumption.

\subsection{Capacity sensitivity}

\begin{figure}[t]
\centering
\includegraphics[width=0.78\textwidth]{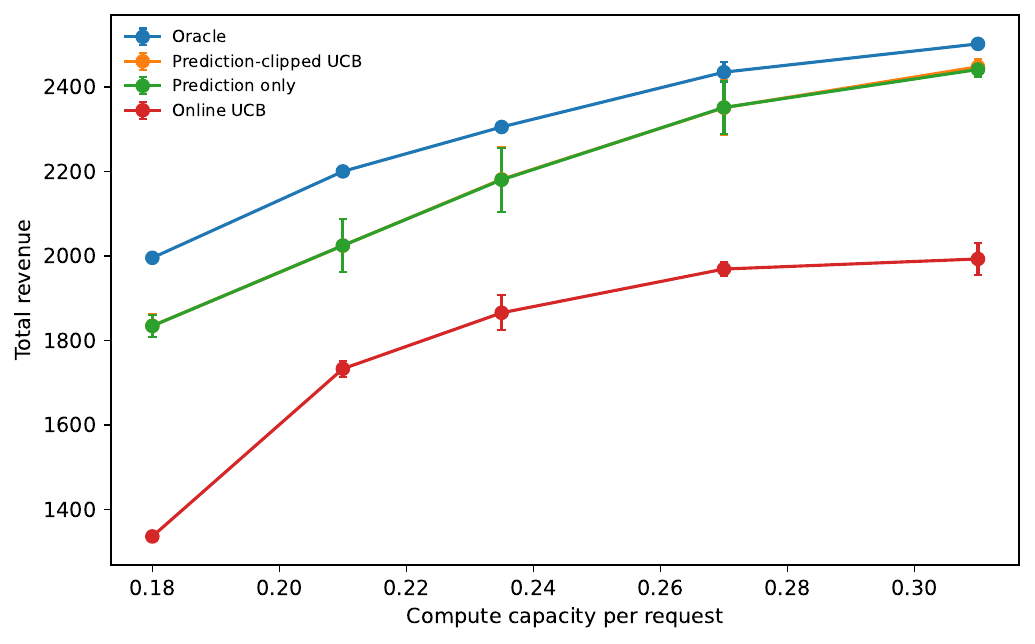}
\caption{Revenue as the compute rate varies; premium capacity is scaled proportionally. The benefit of prediction is largest when capacity is scarce and poor exploration consumes high-opportunity-cost inventory.}
\label{fig:capacity}
\end{figure}

\Cref{fig:capacity} varies compute capacity from 0.18 to 0.31 per request and scales premium capacity at the same ratio used in the main instance. All policies improve with capacity, but online UCB is particularly weak in the scarce regime. A mistaken early offer then has two costs: it produces little revenue and removes capacity from a later high-value request. Prediction-assisted policies avoid much of this loss.

\subsection{Reproducibility}

The source package includes the simulation script, action catalog, repetition-level outcomes, summary table, capacity sweep, mean trajectory, and vector figures. All random seeds are explicit. The code uses the same realized potential outcomes for all policies within each repetition, reducing Monte Carlo noise in policy comparisons. The simulator should be viewed as an executable specification of the model rather than a substitute for an empirical dataset.

\section{Real-Data Evaluation Protocol}\label{sec:realdata}

\subsection{Unit of analysis and product menu}

A credible field study should define the unit before collecting outcomes. One option is an anonymous request session with no carryover; another is an organization-day, which better captures quota interactions. The product menu should be small enough to obtain support for every segment--product cell. A first study could use two approved models, two output caps, and three prices, yielding twelve products plus no offer.

The experiment should record the exact product shown, posted price, exposure timestamp, purchase or authorization decision, model version, prompt length, generated length, latency, cancellations, refunds, and any administrative downgrade. Quality should be measured independently of revenue through task success, verifier scores, safety checks, or blinded pairwise judgments.

\subsection{Identification of demand}

Historical logs are usually insufficient for causal price learning because prices and products were chosen based on latent user value. At least one randomized exploration phase is needed. Within a safe price range, the platform can randomize among menu products using known propensities. Stratification by segment and time reduces variance. The analysis should report overlap; cells with negligible propensity cannot support reliable counterfactual estimates.

A rejection must be distinguished from a system failure. If the platform intended to display an offer but a timeout prevented exposure, coding the outcome as zero demand biases acceptance downward. Similarly, a user who never sees a premium option is not a rejection of that option.

\subsection{Censored token demand}

A token cap censors unconstrained output length. The observed completion length under a short cap cannot be treated as the natural length of the same request under a long cap. Three designs are possible:
\begin{enumerate}[leftmargin=1.6em]
  \item randomize caps and estimate product-specific resource means without extrapolating latent length;
  \item run a consented shadow generation with a longer cap on a small audit sample;
  \item fit a survival or stopping model that treats cap hits as right censoring.
\end{enumerate}
The first is the least assumption-dependent. The second gives richer counterfactual information but consumes compute and raises privacy concerns. The third is efficient only if its model is credible.

\subsection{Constructing the prediction radius}

The offline predictor should be trained on a temporally earlier sample and calibrated on a held-out interval that resembles launch traffic. The reported radius must cover both reward and every resource coordinate. Since the algorithm needs simultaneous validity over many cells, a naive per-cell 95\% interval is insufficient. Bootstrap maxima, conformal calibration under an appropriate exchangeability assumption, or conservative union bounds are possible. The chosen method and its failure rate should be preregistered.

Prediction intervals should be versioned. A model update, price change outside the training range, or new traffic source can invalidate the radius. Empty online--prior intersections provide an operational alarm but do not replace prospective validation.

\subsection{Primary metrics}

The primary economic metric is net revenue or contribution margin, not gross price collected. Inference cost that is already represented as a scarce resource should not be subtracted twice. Capacity utilization should be reported by coordinate, along with the number of meter overrides, cap hits, no-offer decisions, and resource discarded at the horizon.

Quality and user-protection metrics should be co-primary constraints: task success, severe-error rate, latency tail, refund rate, and segment-level service disparity. A policy that raises revenue by degrading a protected or vulnerable user group is not acceptable merely because the resource LP is feasible.

\subsection{Baselines and ablations}

At minimum, the study should compare: a fixed incumbent menu; myopic predicted revenue; prediction-only shadow pricing; online learning without a prior; \PCUCB; and an oracle replay using held-out estimates, labeled clearly as nondeployable. Ablations should remove the reservation meter, replace resource UCBs by point estimates, vary the prediction radius, and vary the product grid. These comparisons isolate whether gains come from price learning, resource pacing, or simply offering a larger menu.

\subsection{Statistical analysis}

Policies should be evaluated on common randomized blocks or through an online switchback design so that traffic composition is comparable. Report cluster-robust uncertainty at the randomization unit. A sequential experiment needs always-valid inference or a prespecified stopping rule. Because capacity creates interference across requests, standard independent-request standard errors can be anti-conservative; organization-day or server-pool-day blocks may be more appropriate.

\section{Extensions}\label{sec:extensions}

\subsection{Continuous prices}

A continuous price can be handled in three ways. The simplest is a fine grid, with approximation error controlled by smoothness of the demand curve. A second uses a parametric purchase model and optimizes the resulting one-dimensional adjusted revenue for each model--cap pair. A third uses a contextual pricing bandit. In all cases, the resource shadow price enters as an additive cost in the objective. The prediction-assisted interval idea applies to the demand-model parameters or directly to adjusted revenue.

\subsection{Nonstationary demand and model versions}

If cell means drift, one can replace full-history statistics by a rolling window or discounting and add a drift allowance to the online intervals. The prior itself may be version-specific. A practical controller should reset cells affected by a model release while retaining information about unchanged prices and segments. The resulting bound would add a variation-dependent term analogous to nonstationary bandit analyses \citep{besbes2015nonstationary}.

\subsection{Repeated and strategic users}

A repeated user may delay purchase in anticipation of a discount, split a long task across low-cap products, or learn which segment label produces a lower price. These effects violate the one-shot demand model. Remedies include commitment to a pricing schedule, organization-level quotas, and incentive-compatible menu design. Personalized prices also raise legal and fairness issues that are outside the present theorem.

\subsection{Queueing-dependent value}

When user value depends on realized latency, demand and resource use are coupled through congestion. One extension adds queue state to the segment and lets shadow prices respond to both cumulative capacity and current backlog. Another separates a slower economic price controller from a fast scheduler that maps the sold service level to a feasible model and batch. Stability then requires queueing analysis in addition to finite-horizon regret.

\subsection{Risk and service-level constraints}

Expected resource constraints do not directly control tail latency or the probability of emergency throttling. Chance constraints, conditional value at risk, or explicit tail-resource coordinates can be included. The reservation envelope already gives a hard upper bound per job, but a useful service-level guarantee may require controlling the joint distribution of many jobs rather than only their sum.

\section{Conclusion}\label{sec:conclusion}

This paper studies an LLM service as a joint pricing and resource-allocation system. Each offer combines a model tier, token cap, and price; accepted requests consume stochastic resources. Prediction-Clipped UCB uses a validated offline interval, online observations, resource shadow prices, and a pathwise reservation meter. The regret analysis shows a smooth transition between trusting an accurate prediction and learning from scratch, while feasibility holds regardless of forecast quality.

The framework suggests a concrete empirical agenda. Build a randomized panel over a small product menu, measure purchase and token use jointly, calibrate simultaneous prediction intervals, and evaluate policies at the workload level rather than request by request. The synthetic results indicate why this matters: ignoring capacity can be disastrous, learning from scratch can be slow, and a prediction that is good but not perfect should be corrected rather than either worshiped or discarded.

\appendix

\section{Proof of the Simultaneous Confidence Lemma}\label{app:proof-confidence}

Fix a cell $j$ and reward coordinate. Conditional on the times at which the cell is selected, its observations are bounded independent draws with mean $r_j$. For any sample size $n\ge1$, Hoeffding's inequality gives
\[
  \Prob\left(
  |\widehat r_{j,n}-r_j|>
  \sqrt{\frac{\log(2J(m+1)T/\delta)}{2n}}
  \right)
  \le \frac{\delta}{J(m+1)T}.
\]
The radius in \eqref{eq:online-radius} is larger by a constant factor. Apply the same argument to each resource coordinate, then take a union bound over $J$ cells, $m+1$ coordinates, and at most $T$ sample sizes. This establishes simultaneous inclusion with probability at least $1-\delta$.

By assumption, the prediction interval also contains the true mean. Therefore the intersection of the online and prediction intervals is nonempty and contains the true mean. The intersection's diameter is no larger than the diameter of either interval, giving
\[
  \diam H_j\le
  \min\{2\varepsilon,2\alpha_t(N_j(t))\}.
\]
Taking the maximum over output coordinates proves \Cref{lem:confidence}.

\section{Proof of the Score Comparison}\label{app:proof-score}

For any cell $j$, both $U^r_j$ and $r_j$ belong to $H^r_j$, so
$|U^r_j-r_j|\le w_j$. Likewise,
$|U^c_{j,i}-c_{j,i}|\le w_j$ for each resource. Hence
\begin{align*}
  \left|
  (U^r_j-\lambda^\top U^c_j)
  -(r_j-\lambda^\top c_j)
  \right|
  &\le |U^r_j-r_j|+
  \sum_{i=1}^m\lambda_i|U^c_{j,i}-c_{j,i}|\\
  &\le (1+\|\lambda\|_1)w_j\\
  &\le(1+\bar\Lambda)w_j,
\end{align*}
where the last line uses the convention that $\bar\Lambda$ bounds the $\ell_1$ norm of the projected price vector. If $\bar\Lambda$ is implemented as a coordinatewise cap, replace it by $m\bar\Lambda$ throughout.

Let $a^\star$ maximize the true score and $\widehat a$ maximize the estimated score. Then
\begin{align*}
  s_{x,\widehat a}(\lambda)
  &\ge \widehat s_{x,\widehat a}(\lambda)
    -(1+\bar\Lambda)w_{x,\widehat a}\\
  &\ge \widehat s_{x,a^\star}(\lambda)
    -(1+\bar\Lambda)w_{x,\widehat a}\\
  &\ge s_{x,a^\star}(\lambda)
    -(1+\bar\Lambda)(w_{x,\widehat a}+w_{x,a^\star}).
\end{align*}
Bounding both widths by their maximum proves the claim. The no-offer action adds the comparison with zero.

\section{Proof of the Confidence-to-Control Theorem}\label{app:proof-generic}

We give the main steps to make the benchmark and error terms transparent. Let $a_t^\star(\lambda_t)$ be a true-score maximizer for segment $X_t$ at price $\lambda_t$, including the no-offer action. Ignoring meter overrides for the moment, \Cref{lem:score} gives
\begin{equation}
  r_{X_t,A_t}-\lambda_t^\top c_{X_t,A_t}
  \ge
  r_{X_t,a_t^\star}-\lambda_t^\top c_{X_t,a_t^\star}
  -2(1+\bar\Lambda)\overline w_t.
  \label{eq:lag-comparison}
\end{equation}
Take conditional expectations. For any feasible fluid decision $y$ at rate $b'$, the segmentwise maximization implies
\begin{align}
  \E[r_{X_t,a_t^\star}-\lambda_t^\top c_{X_t,a_t^\star}\mid\cF_{t-1}]
  \ge
  \frac{\OPT(b')}{T}-\lambda_t^\top b'.
  \label{eq:fluid-lag}
\end{align}
Combining \eqref{eq:lag-comparison}--\eqref{eq:fluid-lag} and rearranging yields
\begin{align}
  \frac{\OPT(b')}{T}-\E[r_{X_t,A_t}]
  \le
  \E[\lambda_t^\top(c_{X_t,A_t}-b')]
  +2(1+\bar\Lambda)\E[\overline w_t].
  \label{eq:perround-gap}
\end{align}

Projected online gradient descent applied to the linear losses
$\ell_t(\lambda)=\lambda^\top(C_t-b')$ gives, for comparator $0$,
\begin{equation}
  \sum_{t=1}^T\lambda_t^\top(C_t-b')
  \le
  \frac{\|\lambda_1\|_2^2}{2\eta}
  +\frac{\eta}{2}\sum_{t=1}^T\|C_t-b'\|_2^2
  \le \frac{\bar\Lambda^2}{2\eta}+\frac{\eta mT}{2},
  \label{eq:ogd}
\end{equation}
up to the choice of projection diameter. With
$\eta=\bar\Lambda/\sqrt{mT}$, the right-hand side is at most a constant times $\bar\Lambda\sqrt{mT}$.

Sum \eqref{eq:perround-gap}, substitute \eqref{eq:ogd}, and add one unit for each meter override. This bounds regret against $\OPT(b')$. Add the buffer sensitivity term \eqref{eq:buffer-loss} and $\delta T$ for the complement of the confidence event. Adjusting constants for the projection geometry gives \eqref{eq:generic-bound}.

A full treatment of stochastic versus mean resource use adds a martingale term
$\sum_t\lambda_t^\top(C_t-c_{X_t,A_t})$. Because prices and resources are bounded, Freedman's inequality controls this term at $O(\bar\Lambda\sqrt{mT\log(1/\delta)})$, which is absorbed by the displayed pacing order and the chosen buffer.

\section{Proof of the Width-Sum Lemma}\label{app:proof-width}

The bound $T\varepsilon$ is immediate. For the sampling branch, let $N_j$ be the total number of visits to cell $j$. Within cell $j$,
\[
  \sum_{n=1}^{N_j}\frac{1}{\sqrt{\max\{1,n-1\}}}
  \le 1+2\sqrt{N_j}.
\]
Multiplying by $\sqrt L$ and summing over cells gives
\[
  \sqrt L\sum_{j=1}^J(1+2\sqrt{N_j})
  \le J\sqrt L+2\sqrt{LJ\sum_jN_j}
  =J\sqrt L+2\sqrt{LJT},
\]
where Cauchy--Schwarz is used in the second inequality. Taking the smaller of the prediction and sampling branches proves \Cref{lem:width-sum}.

\section{Reservation Overrides and Buffer Choice}\label{app:meter}

The reservation meter can override the score maximizer even when cumulative realized consumption is below total capacity, because the next envelope may not fit. Let
$u_{\max,i}=\max_a u_{a,i}$. If, for every resource $i$, the preterminal policy leaves at least $u_{\max,i}$ capacity until the last $O(1)$ rounds, then the number of overrides is bounded by that terminal window. A sufficient high-probability condition is that the buffered target leaves
\[
  T\gamma
  \gtrsim
  \sqrt{T\log(m/\delta)}+\max_i u_{\max,i}+Q_T,
\]
where $Q_T$ bounds the virtual queue generated by the dual update. The first term covers stochastic deviation of realized use from its conditional mean; the second reserves space for one indivisible job; and the third covers tracking error in the pacing controller.

This is intentionally a sufficient rather than tight condition. In the implementation, the meter simply works regardless of the chosen buffer. A small buffer may increase late downgrades or no-offer actions but cannot violate capacity.

\section{Additional Product-Menu Structure}\label{app:menu}

Suppose products $a$ and $a'$ are offered to the same segment and satisfy
\[
  U^r_{x,a}\le U^r_{x,a'},
  \qquad
  U^c_{x,a}\ge U^c_{x,a'}
\]
componentwise. Then $a$ is weakly dominated for every nonnegative price vector and can be removed from the scan. This screening rule uses confidence endpoints. It is safe on the confidence event and can substantially reduce a large discretized menu.

Adjacent prices need not be ordered by revenue because acceptance changes. Nor do larger caps necessarily dominate smaller caps: the larger cap can increase both purchase probability and resource use. The controller should therefore screen on estimated revenue and the full resource vector rather than on product attributes alone.

\section{Complete Synthetic Design}\label{app:simulation}

\subsection{Menu}

The small-model prices are $(0.24,0.36,0.48,0.60)$ and the premium-model prices are $(0.38,0.54,0.70,0.86)$. For each tier, cap $0$ is short and cap $1$ is long. The compute envelope equals a tier factor times a cap factor; the premium envelope is near zero for the small tier and 0.58 for the premium tier.

\subsection{Purchase probability}

Segment base values are $(0.43,0.56,0.68)$. The premium tier adds 0.19. The short cap subtracts 0.035 and the long cap adds 0.075. For product price $p$, purchase probability is
\[
  \sigma\left(\frac{v_{x,a}-p}{0.105}\right),
  \qquad
  \sigma(z)=\frac{1}{1+e^{-z}}.
\]
This creates meaningful but overlapping price choices across segments.

\subsection{Resource use}

Conditional expected compute equals a segment-length term plus a cap adjustment, multiplied by a model-tier factor and clipped below the product envelope. Realized compute adds a mean-zero Gaussian shock and is then clipped to the envelope. Conditional premium use is 0.035 for the small tier and 0.43 for the premium tier, again with a bounded shock. Both resources are zero after a rejection.

\subsection{Prediction error}

A seeded random tensor is augmented by a structured stress direction that increases with premium tier, long cap, and high price. In that direction, predicted revenue is biased upward while compute and premium use are biased downward. The tensor is normalized by its maximum absolute coordinate and multiplied by $\varepsilon$, guaranteeing the advertised sup-norm error bound.

\subsection{Implementation}

Each policy sees the same segment sequence and potential-outcome tensor within a repetition. The score includes a deterministic perturbation of order $10^{-13}$ to break exact ties without affecting reported values. The dual step is 0.045 and prices are projected to $[0,10]$ coordinatewise. Confidence bonuses use a union-bound logarithm over segments, products, output coordinates, and horizon.

\section{Deployment Checklist}\label{app:checklist}

Before a field deployment, the following questions should have explicit answers.

\begin{enumerate}[leftmargin=1.8em]
  \item \textbf{Menu validity:} Are all model--cap--price products technically and legally permissible for every eligible segment?
  \item \textbf{Exposure logging:} Can the system distinguish a displayed offer, user rejection, network failure, and administrative downgrade?
  \item \textbf{Resource envelope:} Is each envelope an enforceable scheduler reservation rather than an empirical percentile mislabeled as a maximum?
  \item \textbf{Prediction coverage:} Was the simultaneous radius validated on traffic representative of the launch horizon?
  \item \textbf{Counterfactual support:} Is there randomized probability on every product that the policy may later consider?
  \item \textbf{Token censoring:} Are cap hits logged and treated as censored demand rather than natural stopping?
  \item \textbf{Quality floor:} Which task-success, safety, and latency constraints can block an economically attractive product?
  \item \textbf{Fairness review:} Which segment features are prohibited from influencing price or service level?
  \item \textbf{Failure mode:} What happens when online and prediction intervals are persistently disjoint?
  \item \textbf{Rollback:} Can the system return immediately to a fixed safe menu if utilization, quality, or user complaints breach a threshold?
\end{enumerate}

\bibliographystyle{plainnat}
\bibliography{references}

@article{ao2026learning,
  title={Learning to price with resource constraints: From full information to machine-learned prices},
  author={Ao, Ruicheng and Jiang, Jiashuo and Simchi-Levi, David},
  journal={Advances in Neural Information Processing Systems},
  volume={38},
  pages={91243--91281},
  year={2026}
}

@article{jiang2025degeneracy,
  title={Degeneracy is ok: Logarithmic regret for network revenue management with indiscrete distributions},
  author={Jiang, Jiashuo and Ma, Will and Zhang, Jiawei},
  journal={Operations Research},
  volume={73},
  number={6},
  pages={3405--3420},
  year={2025},
  publisher={INFORMS}
}

@article{jiang2025tight,
  title={Tight guarantees for multiunit prophet inequalities and online stochastic knapsack},
  author={Jiang, Jiashuo and Ma, Will and Zhang, Jiawei},
  journal={Operations Research},
  volume={73},
  number={3},
  pages={1703--1721},
  year={2025},
  publisher={INFORMS}
}

@article{gallego1994optimal,
  title={Optimal Dynamic Pricing of Inventories with Stochastic Demand over Finite Horizons},
  author={Gallego, Guillermo and van Ryzin, Garrett},
  journal={Management Science},
  volume={40},
  number={8},
  pages={999--1020},
  year={1994}
}

@article{talluri2004revenue,
  title={The Theory and Practice of Revenue Management},
  author={Talluri, Kalyan T. and van Ryzin, Garrett J.},
  journal={Springer International Series in Operations Research and Management Science},
  volume={68},
  year={2004},
  publisher={Springer}
}

@article{besbes2009dynamic,
  title={Dynamic Pricing without Knowing the Demand Function: Risk Bounds and Near-Optimal Algorithms},
  author={Besbes, Omar and Zeevi, Assaf},
  journal={Operations Research},
  volume={57},
  number={6},
  pages={1407--1420},
  year={2009}
}

@article{jasin2014reoptimization,
  title={Reoptimization and Self-Adjusting Price Control for Network Revenue Management},
  author={Jasin, Stefan and Kumar, Sunil},
  journal={Operations Research},
  volume={62},
  number={3},
  pages={588--603},
  year={2014}
}

@article{ferreira2018online,
  title={Online Network Revenue Management Using Thompson Sampling},
  author={Ferreira, Kris Johnson and Simchi-Levi, David and Wang, He},
  journal={Operations Research},
  volume={66},
  number={6},
  pages={1586--1602},
  year={2018}
}

@inproceedings{badanidiyuru2013bandits,
  title={Bandits with Knapsacks},
  author={Badanidiyuru, Ashwinkumar and Kleinberg, Robert and Slivkins, Aleksandrs},
  booktitle={Proceedings of the 54th Annual IEEE Symposium on Foundations of Computer Science},
  pages={207--216},
  year={2013}
}

@inproceedings{agrawal2014bandits,
  title={Bandits with Concave Rewards and Convex Knapsacks},
  author={Agrawal, Shipra and Devanur, Nikhil R.},
  booktitle={Proceedings of the 15th ACM Conference on Economics and Computation},
  pages={989--1006},
  year={2014}
}

@article{balseiro2023best,
  title={The Best of Many Worlds: Dual Mirror Descent for Online Allocation Problems},
  author={Balseiro, Santiago R. and Lu, Haihao and Mirrokni, Vahab},
  journal={Operations Research},
  volume={71},
  number={1},
  pages={101--119},
  year={2023}
}

@article{besbes2015nonstationary,
  title={Non-stationary Stochastic Optimization},
  author={Besbes, Omar and Gur, Yonatan and Zeevi, Assaf},
  journal={Operations Research},
  volume={63},
  number={5},
  pages={1227--1244},
  year={2015}
}

@article{kwon2023vllm,
  title={Efficient Memory Management for Large Language Model Serving with {PagedAttention}},
  author={Kwon, Woosuk and Li, Zhuohan and Zhuang, Siyuan and Sheng, Ying and Zheng, Lianmin and Yu, Cody Hao and Gonzalez, Joseph E. and Zhang, Hao and Stoica, Ion},
  journal={Proceedings of the 29th Symposium on Operating Systems Principles},
  pages={611--626},
  year={2023}
}

@article{yu2022orca,
  title={{Orca}: A Distributed Serving System for Transformer-Based Generative Models},
  author={Yu, Gyeong-In and Jeong, Joo Seong and Kim, Geon-Woo and Kim, Soojeong and Chun, Byung-Gon},
  journal={Proceedings of the 16th USENIX Symposium on Operating Systems Design and Implementation},
  pages={521--538},
  year={2022}
}

@article{sheng2023flexgen,
  title={{FlexGen}: High-Throughput Generative Inference of Large Language Models with a Single {GPU}},
  author={Sheng, Ying and Zheng, Lianmin and Yuan, Binhang and Li, Zhuohan and Ryabinin, Max and Chen, Beidi and Liang, Percy and R{\'e}, Christopher and Stoica, Ion and Zhang, Ce},
  journal={Proceedings of the 40th International Conference on Machine Learning},
  volume={202},
  pages={31094--31116},
  year={2023}
}

@article{chen2023frugalgpt,
  title={{FrugalGPT}: How to Use Large Language Models While Reducing Cost and Improving Performance},
  author={Chen, Lingjiao and Zaharia, Matei and Zou, James},
  journal={arXiv preprint arXiv:2305.05176},
  year={2023}
}

@article{ong2024routellm,
  title={{RouteLLM}: Learning to Route {LLM}s with Preference Data},
  author={Ong, Isaac and Almahairi, Amjad and Wu, Vincent and Chiang, Wei-Lin and Wu, Tianhao and Gonzalez, Joseph E. and Kadous, M. Waleed and Stoica, Ion},
  journal={arXiv preprint arXiv:2406.18665},
  year={2024}
}

@article{hu2024routerbench,
  title={{RouterBench}: A Benchmark for Multi-{LLM} Routing Systems},
  author={Hu, Qitian Jason and Bieker, Jacob and Li, Xiuyu and Jiang, Nan and Keigwin, Benjamin and Ranganath, Gaurav and Keutzer, Kurt and Upadhyay, Shriyash Kaustubh},
  journal={arXiv preprint arXiv:2403.12031},
  year={2024}
}

\end{document}